\documentclass[twocolumn,secnumarabic,amssymb,aps,prd,superscriptaddress]{revtex4}
\usepackage{graphicx,amssymb,amsmath}

\newcommand{\eq}[1]{Eq.~\eqref{#1}}
\newcommand{\fig}[1]{Fig.~\ref{#1}}
\newcommand{\sctn}[1]{\S~\ref{#1}}
\newcommand{\ie}{\textit{i.e.} }
\newcommand{\etc}{\textit{etc.} }
\newcommand{\etal}{\textit{et.~al.} }

\newcommand{\eqnote}[2]{\underset{\textmd{#1}}{\underbrace{#2}}}

\newcommand{\kbt}{k_\textmd{B}T}
\newcommand{\bsm}[1]{B_{\textmd{SM},#1}}
\newcommand{\rsm}{R_\textmd{SM}}
\newcommand{\rsmf}{R_\textmd{SM}^{\prime}}
\newcommand{\vsm}{V_\textmd{SM}^{\prime}}

\newcommand{\nsm}{N_\textmd{SM}}
\newcommand{\rdep}{R_\textmd{dep}}

\newcommand{\phidep}{\phi_\textmd{dep}}
\newcommand{\ndep}{N_\textmd{dep}}
\newcommand{\udep}{u}

\newcommand{\umtm}{\udep_\textmd{MT}}

\newcommand{\uWCA}{U}

\newcommand{\utot}{W}

\newcommand{\vsys}{V_\textmd{sys}}
\newcommand{\effvol}{\nu_\textmd{eff}}
\newcommand{\threebody}{\omega_\textmd{eff}}
\newcommand{\critThree}{\omega_\textmd{eff}^c}
\newcommand{\rg}{R_\textmd{g}}

\newcommand{\nm}{~\textmd{nm}}
\newcommand{\kbp}{~\textmd{kbp}}

\begin{document}

%%%%%%%%%%%%%%%%%%%%%%%%%%%%%%%%%%%%%%%%%%%%%%%%%%%%%%%%%%%%
%%%%%%%%%%%%%%%%%%%%%%%%%%%%%%%%%%%%%%%%%%%%%%%%%%%%%%%%%%%%
%%%%%%%%%%%%%%%%%%%%%%%%%%%%%%%%%%%%%%%%%%%%%%%%%%%%%%%%%%%%
%Prefix
%%%%%%%%%%%%%%%%%%%%%%%%%%%%%%%%%%%%%%%%%%%%%%%%%%%%%%%%%%%%
%%%%%%%%%%%%%%%%%%%%%%%%%%%%%%%%%%%%%%%%%%%%%%%%%%%%%%%%%%%%
%%%%%%%%%%%%%%%%%%%%%%%%%%%%%%%%%%%%%%%%%%%%%%%%%%%%%%%%%%%%

\title{Simulating the Entropic Collapse of Coarse-Grained Chromosomes}

\author{Tyler N. Shendruk}
\email{tyler.shendruk@physics.ox.ac.uk}
\affiliation{The Rudolf Peierls Centre for Theoretical Physics, Department of Physics, Theoretical Physics, University of Oxford, 1 Keble Road, Oxford, OX1 3NP, United Kingdom}
\author{Martin Bertrand}
\affiliation{University of Ottawa, Department of Physics, 150 Louis-Pasteur, Ottawa, ON, K1N 6N5, Canada}
\author{Hendrick W. de~Haan}
\affiliation{University of Ontario Institute of Technology, Faculty of Science, 2000 Simcoe St. North, Oshawa, ON, Canada}
\author{James L. Harden}
\affiliation{University of Ottawa, Department of Physics, 150 Louis-Pasteur, Ottawa, ON, K1N 6N5, Canada}
\author{Gary W. Slater}
\email{gslater@uottawa.ca}
\affiliation{University of Ottawa, Department of Physics, 150 Louis-Pasteur, Ottawa, ON, K1N 6N5, Canada}
\date{\today}

\begin{abstract}
Depletion forces play a role in the compaction and de-compation of chromosomal material in simple cells but has remained debatable whether they are sufficient to account for chromosomal collapse. 
We present coarse-grained molecular dynamics simulations, which reveal that depletion-induced attraction is sufficient to cause the collapse of a flexible chain of large structural monomers immersed in a bath of smaller depletants. 
These simulations use an explicit coarse-grained computational model that treats both the supercoiled DNA structural monomers and the smaller protein crowding agents as combinatorial, truncated Lennard-Jones spheres. 
By presenting a simple theoretical model, we quantitatively cast the action of depletants on supercoiled bacterial DNA as an effective solvent quality. 
The rapid collapse of the simulated flexible chromosome at the predicted volume fraction of depletants is a continuous phase transition. 
Additional physical effects to such simple chromosome models, such as enthalpic interactions between structural monomers or chain rigidity, are required if the collapse is to be a first-order phase transition. 
\end{abstract}

\maketitle

%%%%%%%%%%%%%%%%%%%%%%%%%%%%%%%%%%%%%%%%%%%%%%%%%%%%%%%%%%%%
%%%%%%%%%%%%%%%%%%%%%%%%%%%%%%%%%%%%%%%%%%%%%%%%%%%%%%%%%%%%
%%%%%%%%%%%%%%%%%%%%%%%%%%%%%%%%%%%%%%%%%%%%%%%%%%%%%%%%%%%%
%Introduction
%%%%%%%%%%%%%%%%%%%%%%%%%%%%%%%%%%%%%%%%%%%%%%%%%%%%%%%%%%%%
%%%%%%%%%%%%%%%%%%%%%%%%%%%%%%%%%%%%%%%%%%%%%%%%%%%%%%%%%%%%
%%%%%%%%%%%%%%%%%%%%%%%%%%%%%%%%%%%%%%%%%%%%%%%%%%%%%%%%%%%%
\section{Introduction}
The physical organization of chromosomes plays an essential role during cell division and in determining gene activity. 
While eukaryote cells have extensive cellular machinary that is known to be dedicated to anaphase separation of the chromatids~\cite{luijsterburg08}, bacteria are significantly simpler and entropic contributions are suspected to play an important role. 
Entropic repulsion between replicated daughter strands within prokaryotes can be sufficient for segregation: 
Excluded volume interactions between the chain segments determine whether they will remain mixed or spontaneously separate within the nucleoid~\cite{jun06,arnold07,fan07,jun10,jun10b,jung12,jung12b,minina14}. 
Thus, under high confinement conditions, entropy can drive DNA to recede to opposite cellular poles. 

It has been suggested that entropic considerations are essential (if not sufficient), not only for segregating daughter strands but also for the compaction and de-compaction of bacterial chromosomes~\cite{walter95,zimmerman96}. 
Molecular crowding by surrounding cytoplasmic proteins enacts a depletion attraction between components of the chromosome and it has been proposed that these may be strong enough to cause a phase transition from a swollen conformation to a collapsed globular state. 
Condensation of DNA by macromolecular crowding effects has been known for many years~\cite{lerman71,minton81,minton83} and neutral PEG polymers acting as depletants can significantly reduce the radius of gyration of large macromolecules~\cite{kojima06}. 
Since the volume fraction of cytoplasmic proteins is approximately $20\%$ in \textit{Escherichia coli} cells, depletion forces are non-negligible~\cite{zimmerman91,woldringh99}. 
Experimental observations of macromolecules within eukaryotic nuclei suggest that these crowding effects influence interactions within cells~\cite{cunha01,richter07,hancock08}. 
However, questions still remain about the nature of the transition from swollen to collapsed state in simple, prokaryotic chromosomes. 
Indeed, it has yet to be confirmed whether or not entropic forces are sufficient to account for chromosomal collapse, though recent experimental work has suggested that a reported first-order coil-globule collapse of \textit{E. coli} chromosomes occurs in depletant baths of PEG at a volume fraction of $11-13\%$ in microchannels, based on the apparent co-existence of swollen coils and collapsed globules~\cite{pelletier12}. 
Odijk theory for the compactification of supercoiled DNA~\cite{odijk98} was modified to account for this reported first-order coil-to-globule phase transition~\cite{pelletier12}. 
This is in agreement with the observed first-order transition of T4DNA molecules in concentrations of spermidine~\cite{yoshikawa96} and nucleoids in solutions of PEG and MgCl$_2$ or spermidine~\cite{zimmerman06}, though it is important to note that the trivalent nature of spermidine introduces complications to crowding, as the depletion-induced compaction process may act to pre-position DNA segments for the action of multivalent ions and three-body DNA-protein interactions. 

Due to unavoidable complications in such experimental systems, computational simulations are needed to verify whether or not non-specific depletion forces are sufficient to cause the collapse of bacterial chromosomal DNA. 
In this manuscript, we study the effect of depletants on the conformation of a chain of idealized DNA structural monomers. 
We do so using generic computational methods in order to consider a simplified system. 

By neglecting much of the biological complications of prokaryotic chromosomal material crowded by proteins with specific protein-DNA interactions in a confined space, our simulations are able to  explicitly test the hypothesis that depletant-induced attraction can be sufficient to cause the collapse of bacterial chromosomes from a swollen state to a globular state~\cite{pelletier12}. 
Our results demonstrate that the presence of smaller depletants is indeed sufficient to cause the collapse of a chain of freely joined spherical structural monomers (as a model chromosome) in the absence of confinement. 
We stress that our simulations show that this model is sufficient to account for a continuous coil-to-globule transition but does not produce a first-order phase transition, as suggested by the \textit{in vitro} experiments of Pelletier \etal~\cite{pelletier12}. 
While it should be noted that these experiments were not performed under physiological conditions but rather in microfluidic environments, the presumption that structural monomers are akin to freely jointed excluded-volume spheres in an inert bath of depletants requires reconsideration in light of our results. 
Additional complications must be accounted for. 
Physical forces such as electrostatic, dielectrophoretic, hydrophobic and chain stiffness may all play a role. 
In addition, there is strong evidence that DNA supercoiling facilitates compactification~\cite{tolstorukov05} and it is well-known that RNA polymerase and several DNA-associated proteins are localized within nucleoids. 
These include both small nucleoid-associated (or histone-like) proteins~\cite{thanbichler06,luijsterburg06} and structural-maintenance-of-chromosomes (SMC) complexes~\cite{hirano05,hirano06}. 
Such specific interactions mediated by nucleoid-associated proteins are likely crucial for the formation of mesoscale chromosomal structure~\cite{minsky04,stavans06}.
While our coarse-grained approach is not able to determine which of these mechanisms change the nature of the compaction process, we present strong evidence that entropic effects cannot be solely responsible and that specific biological interactions must be included in coarse-grained models of chromosome condensation.
However, depletion effects undoubtedly occur in crowded systems such as in the cell, and it seems likely that depletion-induced compactification, even if continuous in nature, facilitates the action of other enthalpic interactions between DNA and proteins. 

Furthermore, this work quantitatively maps the collapse-behaviour onto the language of solvent quality, through a simplified Flory theory.
We demonstrate that conceptually simple models for depletion-induced interactions accurately predict the effective solvent quality for sufficiently large ratios of DNA structural monomers to protein depletants. 
In this way, theoretical curves of the coil-globule order parameter exhibiting the same continuous coil-globule collapse agree well with the coarse-grained simulations. 
Expressing the effect of the depletants as an effective solvent quality suggests that a first-order coil-globule collapse requires smaller three-body interactions, which may be possible by including additional physical effects such as enthalpic protein interactions, chain rigidity or prescribed interactions between structural monomers. 

%%%%%%%%%%%%%%%%%%%%%%%%%%%%%%%%%%%%%%%%%%%%%%%%%%%%%%%%%%%%
%%%%%%%%%%%%%%%%%%%%%%%%%%%%%%%%%%%%%%%%%%%%%%%%%%%%%%%%%%%%
%%%%%%%%%%%%%%%%%%%%%%%%%%%%%%%%%%%%%%%%%%%%%%%%%%%%%%%%%%%%
%Methods
%%%%%%%%%%%%%%%%%%%%%%%%%%%%%%%%%%%%%%%%%%%%%%%%%%%%%%%%%%%%
%%%%%%%%%%%%%%%%%%%%%%%%%%%%%%%%%%%%%%%%%%%%%%%%%%%%%%%%%%%%
%%%%%%%%%%%%%%%%%%%%%%%%%%%%%%%%%%%%%%%%%%%%%%%%%%%%%%%%%%%%
\section{Simulation Methods}
\begin{figure}
 \begin{center}
  \includegraphics[width=0.45\textwidth]{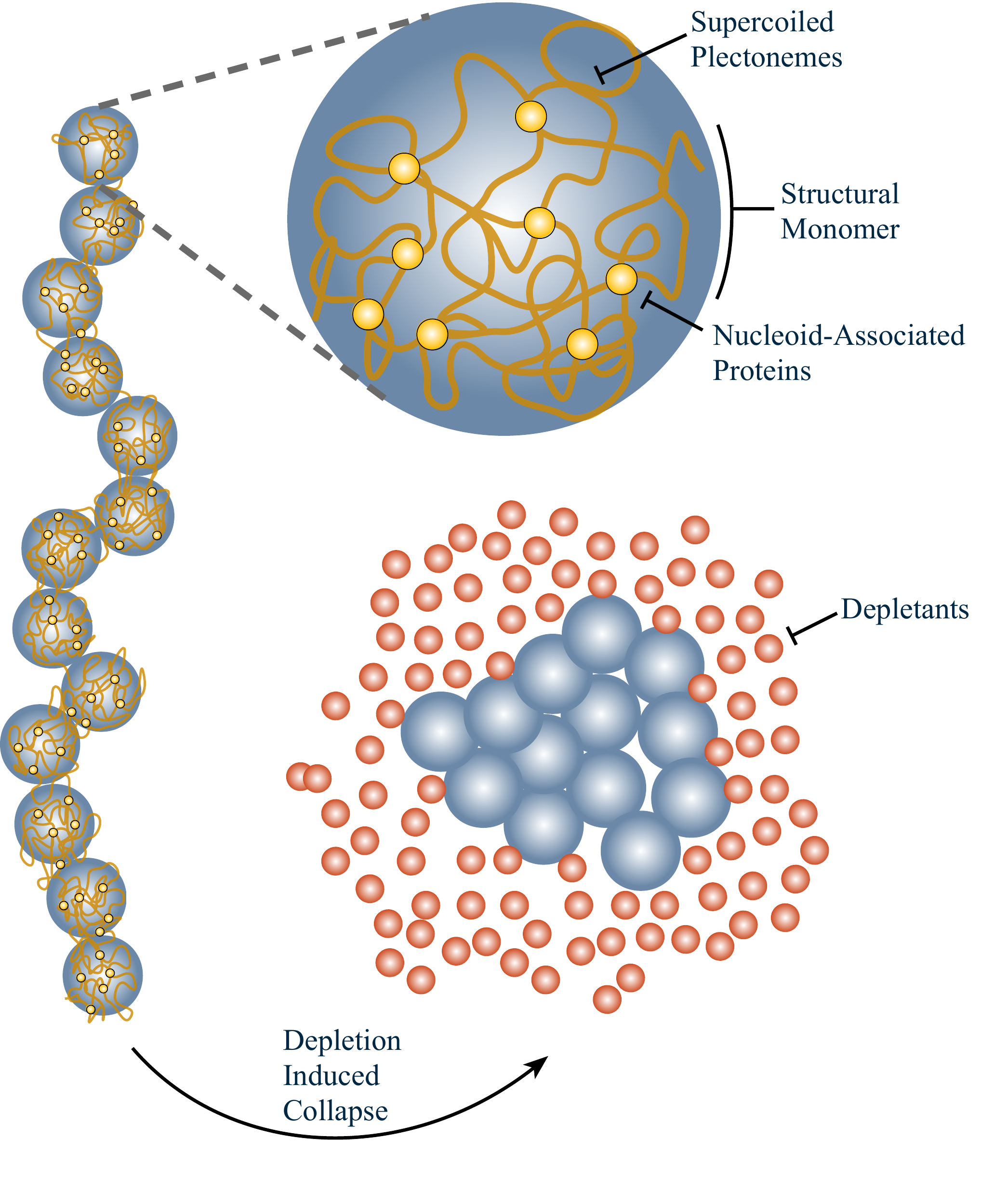}
  \caption{The coarse-grained model of bacterial chromosomal DNA after Jun and Wright~\cite{jun10}. 
  %(a) 
  Structural monomers of supercoiled plectonemes of DNA are locally stabilized to form a crosslinked gel by various nucleoid-associated proteins. 
  %(b) 
  The chromosome is considered to be a linear chain of structural monomers. 
  %(c) 
  Surrounding proteins act as molecular crowding agents that can lead to collapse to a condensed state.}
  \label{fig:SM}
 \end{center}
\end{figure}

Throughout this study, we will consider a coarse-grained model of bacterial chromosomes. 
This imparts simplicity but also allows our conclusions to be generally applicable to the action of depletants on large biomolecules. 
Here, each chromosome is viewed as a linear, freely jointed chain of DNA structural monomers (SMs). 
Structural monomers have also been referred to as ``structural units''~\cite{jun10,pelletier12} and ``compacted domains''~\cite{fan07} in the literature. 
Each structural monomer is a distinct topological domain of supercoiled plectonemes~\cite{postow04} that is stabilized by crosslinking via nucleoid-associated proteins (\fig{fig:SM})~\cite{stavans06,woldringh06,jun10}. 
It is likely that many different SMC proteins~\cite{graumann01,dame05,stavans06,pelletier12} crosslink each structural monomer. 
The chromosome separates into a dense protein-poor nucleoid of structural monomers and an exterior of protein-rich cytoplasm~\cite{valkenburg84,woldringh10}. 
Each structural monomer is thus modelled as a spherical monomer of radius $\rsm$ (volume $V_\textmd{SM}\sim\rsm^3$) that sterically excludes all non-nucleoid-associated proteins from entering its interior. 

Both the structural monomers and the cytoplasmic protein depletants are modelled as inert, hard spherical particles. 
Our coarse-grained simulations consist of representing DNA structural monomers and many smaller depletants as hard spheres diffusing within an implicit Langevin solvent. 
The solvent is included solely in a statistical manner by replacing the explicit fluid with drag and Brownian forces.
The hard spheres are modelled as truncated Lennard-Jones beads, which are purely repulsive radial combinatorial Weeks-Chandler-Andersen (WCA) potentials
\begin{small}\begin{align}
 \uWCA &=
 \begin{cases}
   4\varepsilon_{ij}\left[ \left(\frac{\sigma_{ij}}{r_{ij}}\right)^{12} - \left(\frac{\sigma_{ij}}{r_{ij}}\right)^{6}\right] + \varepsilon_{ij} &r_{ij} < r^\textmd{cut}_{ij} \\
   0 & r_{ij} \geq r^\textmd{cut}_{ij},
  \end{cases}
 \label{eq:WCA}
\end{align}\end{small}
where $r_{ij}$ is the centre-to-centre separation between particles $i$ and $j$, $\varepsilon_{ij}$ is the depth of the potential well, $\sigma_{ij}=R_{i} + R_{j}$ is the effective size of the pair of particles, and $r^\textmd{cut}_{ij} = 2^{1/6}\sigma_{ij}$ is the cutoff radius. 
The energy and length scale units of these simulations are denoted $\varepsilon$ and $\sigma$. 
We set $\varepsilon_{ij}=1 \varepsilon = \kbt$ regardless of whether $i$ and $j$ are both colloids, both depletants or a colloid/depletant pair, reflecting the nonspecific nature of such systems. 
Depletants are assigned a size of $\rdep=0.5\sigma$ and structural monomers of sizes $\rsm = \left\{1.5,2,2.5\right\}\sigma$ are considered throughout. 
Although $\rsm$ and $\rdep$ are the sizes used in \eq{eq:WCA}, the statistical effective size of the structural monomers must determined from the second virial coefficient. 
One can numerically calculate that, in the absence of depletants, the structural monomers have a statistical effective size of $\rsmf=1.0174\rsm$~\cite{part1}. 

It is important to note that the detailed choices made in the WCA potential are essential. 
When the system of structural monomers and depletants interact via the combinatorial-WCA model, the depletant-induced pair potential between structural monomers is deeper than for hard spheres. 
This is because the WCA repulsion potential rises continuously, rather than discontinuously jumping to infinity, allowing the centre-to-centre separation to be less than $2\rsm$ and decreasing the excluded volume that the depletants cannot occupy. 
This in turn increases the osmotic pressure and deepens the attractive well depth. 
Elsewhere~\cite{part1}, we have shown that the net pair potential can be predicted by summing the combinatorial-WCA interaction and the depletant-induced component as modelled by Morphometric Thermodynamics (MT)~\cite{oettel09,botan09,ashton11}. 
Here, we use the Rosenfeld functionals in the MT model~\cite{rosenfeld89,botan09} and allow the resulting pair potential to extend to smaller separations than the contact point between two monomers and employ the statistical size of the structural monomers $\rsmf$. 
From this model, the second virial coefficient can be calculated for an ensemble of structural monomers in an implicit depletant bath~\cite{part1}.  
The approximation is a correction to the constant hard-sphere value $4\vsm$, while the third virial coefficient is assumed to be adequately approximated as the hard-sphere value of $10{\vsm}^2$ because depletant-induced interactions are non-additive. 

This coarse-grained simulation method and the corresponding theoretical model will now be used to study the behaviour of chromosomes in a bath of inert depletant proteins and test whether depletant-induced attraction is sufficient to cause the collapse of bacterial chromosomes to a globular state. 
Since a typical bacteria chromosome consists of $\sim4.6$ million base pairs (\textit{E. coli}), each structural monomer contains $\sim300\kbp$. 
The physical size of structural monomers has been estimated as low as $80\nm$~\cite{romantsov07} and as high as $440\nm$~\cite{pelletier12}, with many estimates falling between~\cite{jun10,cunha01,woldringh02}. 
Experimental measurements~\cite{pelletier12} suggest that the lower bound estimate for the number of structural monomers is $\nsm\approx16$ and in this study we simulate chains of $\nsm=15$ monomers. 
In the coarse-grained simulations, structural monomers are polymerized into a chain via FENE springs~\cite{slater09}. 

We consider three ratios between the size of the structural monomers and of the depletants: $\rsm/\rdep=\left\{3,4,5\right\}$. 
Typical proteins would produce more realistic size ratios in the range $\rsm/\rdep\approx20-100$. 
However, the division of time scales required to produce sufficient statistics for the structural monomers and yet resolve interactions between depletants becomes computationally severe as $\rsm/\rdep$ becomes larger and such ratios are not computationally feasible. 
As we shall see, depletion effects are weaker for near-unity ratios (though not as weak as previously thought~\cite{zaccone12}) and so coil-globule collapse at such small size ratios implies that collapse will occur at larger ratios, though we leave confirmation to future studies, which may utilize simulations in which depletion effects are included implicitly. 

Every simulation data point reported here is an average of three simulations that each ran for $5\times10^7$ time-steps after a short warm-up period. 
In order to minimize finite size effects, which are generally known to play an important role in phase transitions, periodic boundary conditions were implemented on control volumes of $\vsys = \left\{ 30^3, 40 ^3, 60^3 \right\} \sigma^3$ for $\rsm = \left\{1.5,2,2.5\right\}\sigma$, respectively. 
This required $\ndep$ as large as $294801$ in order to achieve a maximum volume fraction of $\phidep = 0.45$. 
It should be stressed that chromosomal systems are naturally confined and depletion effects in the presence of confining walls may play a significant role in determining \textit{in vivo} conformation~\cite{mondal11}. 

%%%%%%%%%%%%%%%%%%%%%%%%%%%%%%%%%%%%%%%%%%%%%%%%%%%%%%%%%%%%
%%%%%%%%%%%%%%%%%%%%%%%%%%%%%%%%%%%%%%%%%%%%%%%%%%%%%%%%%%%%
%%%%%%%%%%%%%%%%%%%%%%%%%%%%%%%%%%%%%%%%%%%%%%%%%%%%%%%%%%%%
%Polymer Simulations
%%%%%%%%%%%%%%%%%%%%%%%%%%%%%%%%%%%%%%%%%%%%%%%%%%%%%%%%%%%%
%%%%%%%%%%%%%%%%%%%%%%%%%%%%%%%%%%%%%%%%%%%%%%%%%%%%%%%%%%%%
%%%%%%%%%%%%%%%%%%%%%%%%%%%%%%%%%%%%%%%%%%%%%%%%%%%%%%%%%%%%
\section{Depletion-Induced Polymer Collapse}

\subsection{From Swollen to Collapsed State}

\begin{figure}[tb]
 \begin{center}
  \includegraphics[width=0.49\textwidth]{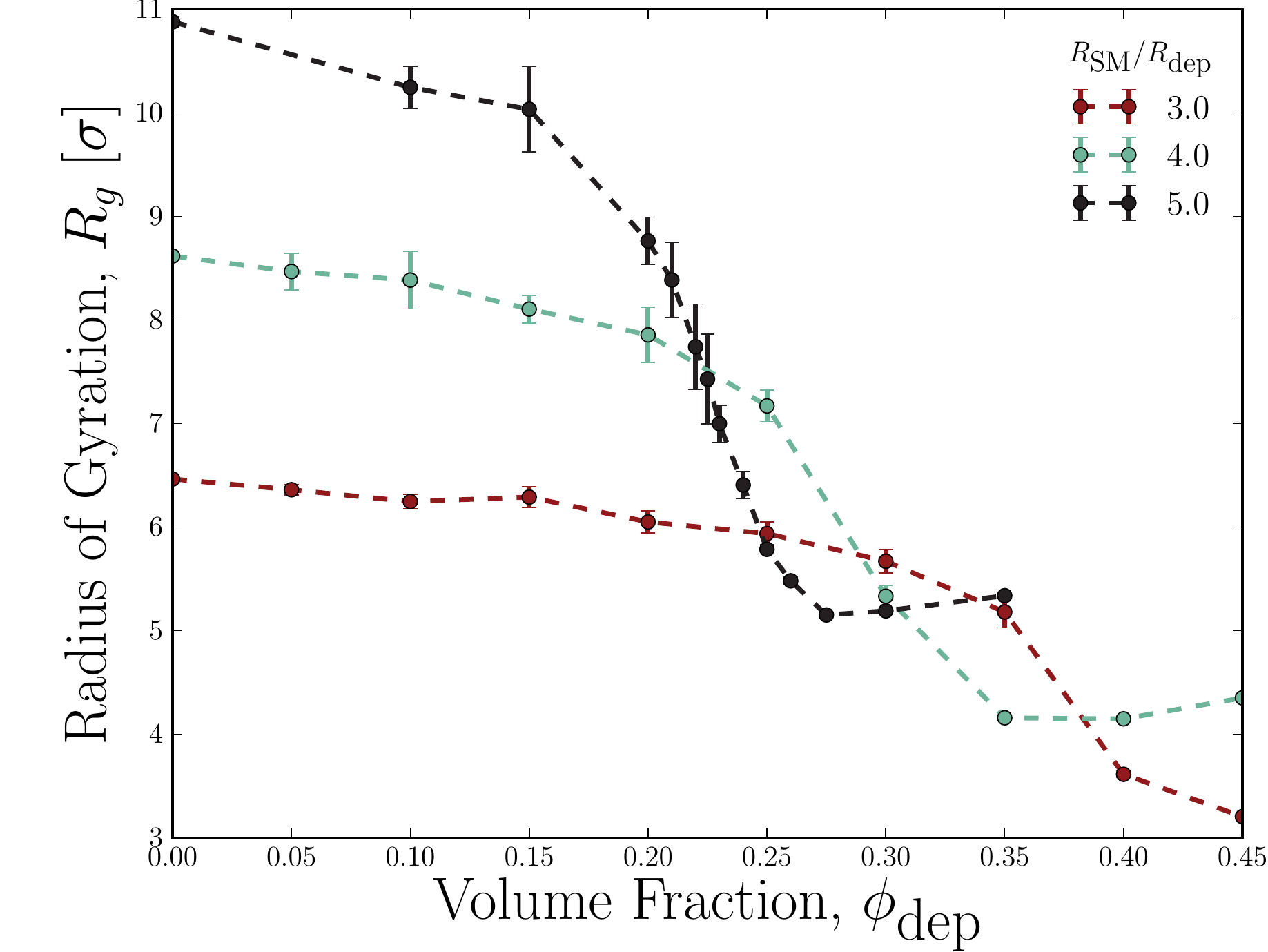}
  \caption{
  The radii of gyration of $\nsm=15$ chains of combinatorial-WCA structural monomers as a function of depletant volume fraction, $\phidep$. The observed transition to a collapsed state varies as a function of the size ratio $\rsm/\rdep$.
  }
  \label{fig:softCollapse}
 \end{center}
\end{figure}

For model chromosome chains in the absence of depletant proteins and confinement, excluded volume interactions between structural monomers swell the polymer to a radius of gyration greater than the ideal value of $\rg\approx6^{-1/2}\rsmf\nsm^{1/2}$. 
Qualitatively, we expect the free solution radius of gyration $\rg$ to decrease as the volume fraction of depletants $\phidep$ is increased because the entropic forces increase. 
Indeed, \fig{fig:softCollapse} demonstrates that as the number of depletants is increased, the simulated radius of gyration decreases. 
In fact, the radius of gyration collapses from its large (swollen state) value at $\phidep=0$ to a much smaller, compact state (globular state) when combinatorial-WCA simulations are performed (\fig{fig:softCollapse}). 

Since $\rsm/\rdep=3$ is the smallest ratio of sizes considered, the chain's radius of gyration falls from a swollen to a collapsed state at a high volume fraction of depletants ($0.35\lesssim\phidep\lesssim0.4$), which is approaching the highest volume fractions investigated using explicit simulations. 
Let us define the critical depletant volume fraction for the simulations $\phidep^{*_\textmd{cWCA}}$ to be the point when the chain reaches its minimum globular radius of gyration. 
At the larger size ratio of $\rsm/\rdep=4$, the drop occurs at lower volume fractions and the chain collapses to a compact globular state by $\phidep^{*_\textmd{cWCA}}=0.35$. 
At the largest size ratio $\rsm/\rdep=5$, the drop occurs between $0.15\lesssim\phidep\lesssim0.275$ and has a measured critical volume fraction of $\phidep^{*_\textmd{cWCA}}=0.275$. 
For larger size ratios (as expected when cytoplasmic proteins act as depletants on the chromosome), the critical point is expected to reside at even lower volume fractions. 

The $\rsm/\rdep=5$ curve appears to have a small rise after collapse (\fig{fig:softCollapse}). 
Indeed, at the highest densities of depletants, the chain rapidly collapses into metastable conformational states that require uncommonly large fluctuations in order to escape and find the global minimum in free energy. 
In most cases, our simulations could not reach the minimum free energy state over the duration of our simulations. 

Simulations were also performed in which the structural monomers were modelled using the steeper shifted-WCA potential. 
Since the depletion-induced well depths produced by the shifted-WCA potential are far shallower, no collapse to a globular state was observed (not shown). 
A small decrease did occur for $\rsm/\rdep=5$ but by the highest volume fractions accessible to simulations ($\phidep\approx0.4$) a substantial drop like those observed for the combinatorial-WCA model was not observed. 
This is because the combinatorial-WCA model produces deeper well depths, causing the coil-globule collapse to occur at lower (computationally accessible) volume fractions. 

\begin{figure}[tb]
 \begin{center}
  \includegraphics[width=0.49\textwidth]{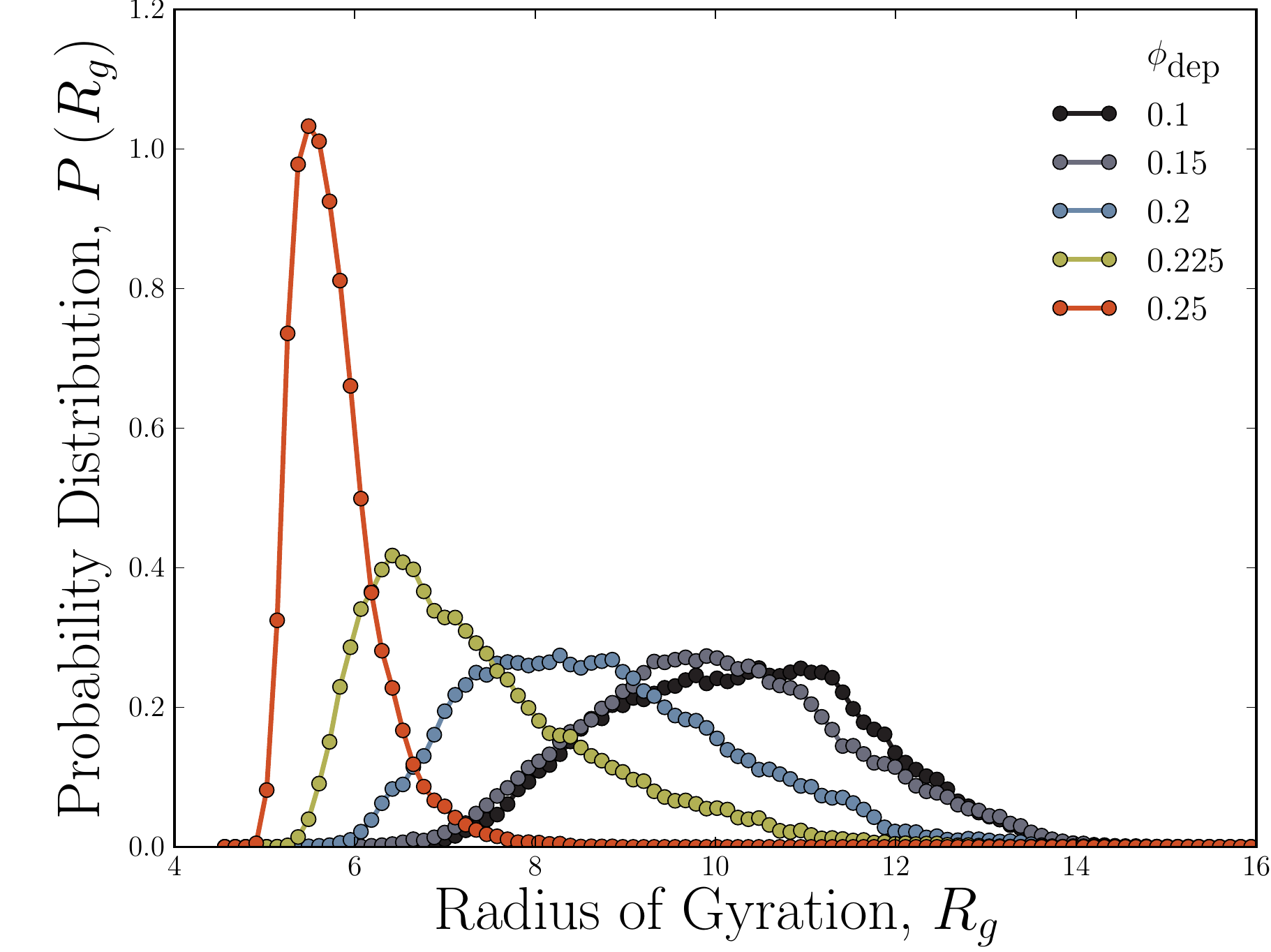}
  \caption{Distributions of radii of gyration $\rg$ for the size ratio of $\rsm/\rdep=5$ and various volume fractions of depletants. 
  The volume fractions shown are in the transition region, near the critical volume fraction of depletants $\phidep^*$ but the probability distributions remain unimodal. 
  There appears to be no coexistence of both swollen and collapsed coils. 
  }
  \label{fig:dists}
 \end{center}
\end{figure}

The collapses in \fig{fig:softCollapse} are evidently not discontinuous. 
In particular, the uncertainty on the radii of gyration during the collapse is not comparable to the drop. 
The probability distribution of the radius of gyration $P\left(\rg\right)$ in the absence of depletants is broad since fluctuations of a polymer in good solvent are large. 
\fig{fig:dists} shows the radius of gyration for $\rsm/\rdep=5$ as a function of depletant volume fraction in free solution. 
As the volume fraction of depletants is increased, both the mean and the most likely value decrease and at the highest volume fractions of depletants the probability distribution becomes sharply peaked about the collapsed state (\fig{fig:dists}). 
The chains have collapsed to a globular state and do not fluctuate significantly. 
However, throughout the transition, coexistence of the swollen state and the globular state is not observed to occur. 
Each of these probability distributions is unimodal, containing only a single maximum and at no point during the transition can we identify co-existence between swollen and globular states. 
Thus, the transition is a continuous collapse as is expected for a freely jointed polymer~\cite{grosberg94}, but this finding is in disagreement with the experimental report of Pelletier \etal~\cite{pelletier12}. 

%%%%%%%%%%%%%%%%%%%%%%%%%%%%%%%%%%%%%%%%%%%%%%%%%%%%%%%%%%%%
%%%%%%%%%%%%%%%%%%%%%%%%%%%%%%%%%%%%%%%%%%%%%%%%%%%%%%%%%%%%
%%%%%%%%%%%%%%%%%%%%%%%%%%%%%%%%%%%%%%%%%%%%%%%%%%%%%%%%%%%%
%Polymer Theory
%%%%%%%%%%%%%%%%%%%%%%%%%%%%%%%%%%%%%%%%%%%%%%%%%%%%%%%%%%%%
%%%%%%%%%%%%%%%%%%%%%%%%%%%%%%%%%%%%%%%%%%%%%%%%%%%%%%%%%%%%
%%%%%%%%%%%%%%%%%%%%%%%%%%%%%%%%%%%%%%%%%%%%%%%%%%%%%%%%%%%%
\section{Chromosome-Depletant Flory Theory} \label{polyTheory}
We have observed the coil-globule collapse of a highly idealized model of a prokaryotic chromosome in free space by the entropic action of depletants alone (\fig{fig:softCollapse}). 
We now wish to discuss this coil-globule collapse within a simple theoretical framework.  
To do so, we construct a classic Flory theory for chromosomal DNA in an implicit bath of protein depletants. 
More detailed theories build a free energy for the entire system of chromosomal material and cytoplasmic proteins~\cite{odijk98}. 
Here, we take a far more simplistic approach: 
rather than explicitly accounting for the protein depletants, we only implicitly include them in an effective interaction free energy between structural monomers through an effective volume $\effvol$ and a three-body interaction coefficient $\threebody$. 
Implicitly including the effects of depletants in the free energy has previously proven useful for understanding polymer collapse in an ensemble of crowders that have the same size as the monomers~\cite{zaccone12}. 

\subsection{Free Energy and the Radius of Gyration}

When the depletants are included implicitly, the free energy of the system is entirely due to the chromosome chain alone. 
The free energy of the system then has only two terms $F = F_{ent} + F_{int}$. 
The first term is an entropic-spring free energy cost due to connectivity
\begin{align}
  \frac{F_{ent}}{\kbt} &\simeq \frac{R_g^2}{\nsm{\rsmf}^2} + \frac{\nsm{\rsmf}^2}{R_g^2},
  \label{ent}
\end{align}
where all near unity numerical coefficients are omitted throughout this discussion. 
This is a common interpolation between the free energy of a swollen state ($F_{ent}\propto R_g^2$) and a collapsed state ($F_{ent}\propto R_g^{-2}$)~\cite{grosberg94,grosberg97}. 
The second free energy term is an effective interaction free energy that can be written as an expansion 
\begin{align}
  \frac{F_{int}}{\kbt} &\simeq \left[ \left(\frac{\nsm^2}{R_g^3}\right) \effvol + \left(\frac{\nsm^3}{R_g^6}\right) \threebody + \ldots \right] ,
  \label{int}
\end{align}
where $\effvol$ is the effective excluded volume, $\threebody$ is the three-body interaction coefficient, \etc 

This is a simplistic but robust and general way to approach a generic polymer. 
Minimizing $F$ with respect to $R_g$ gives an expression for the ratio $a\equiv\left(R_g/\rsmf\right) \nsm^{-1/2}$ of the form
\begin{align}
 \eqnote{swelling}{a^5} - \eqnote{compr.}{a} &\simeq 
   \eqnote{2-body int.}{ \left(\frac{\effvol}{{\rsmf}^3}\right)\nsm^{1/2}}
   + \eqnote{3-body int.}{\left(\frac{\threebody}{{\rsmf}^6}\right)a^{-3}}. 
   \label{rg}
\end{align}
Also bear in mind that $\rsmf$ is the effective radius of the structural monomers as modelled by the WCA potential in the absence of depletants, while $\effvol$ is the effective volume of the structural monomers in the presence of a depletant bath. 
Calculating $\effvol$ will be discussed in the upcoming section (\sctn{sctn:phaseTrans}).

Our strategy is to consider the limiting cases of \eq{rg} and model the interaction coefficients of the polymer chain as the virial coefficients of an ensemble of free (\ie unconnected) structural monomers interacting in a bath of small particles that are included only implicitly via depletation forces \ie through $ \effvol$ and $\threebody$. 
Quantitatively, the effective volume of the structural monomers $\effvol$ controls the solvent quality through the usual definition 
\begin{align}
 \chi &= \frac{1}{2}-\frac{\effvol}{4\vsm}.
 \label{solventQuality}
\end{align}
Therefore, we propose to discuss the chromosome-depletants system in terms of an effective solvent quality due to the depletion forces~\cite{hancock08}. 
Each solvent regime is an idealization in which all but two of the terms in \eq{rg} are considered to be insignificant and neglected. 

%%%%%%%%%%%%%%%%%%%%%%%%%%%%%%%%%%%%%%%%%%%%%%%%%%%%%%%%%%%%
%%%%%%%%%%%%%%%%%%%%%%%%%%%%%%%%%%%%%%%%%%%%%%%%%%%%%%%%%%%%
%%%%%%%%%%%%%%%%%%%%%%%%%%%%%%%%%%%%%%%%%%%%%%%%%%%%%%%%%%%%
%Solvent Theory
%%%%%%%%%%%%%%%%%%%%%%%%%%%%%%%%%%%%%%%%%%%%%%%%%%%%%%%%%%%%
%%%%%%%%%%%%%%%%%%%%%%%%%%%%%%%%%%%%%%%%%%%%%%%%%%%%%%%%%%%%
%%%%%%%%%%%%%%%%%%%%%%%%%%%%%%%%%%%%%%%%%%%%%%%%%%%%%%%%%%%%
\begin{subequations}\label{quality:eq}
\paragraph{Good Solvent}
\label{goodSolvent}
If the coil is in a swollen state then three-body interactions are rare and the compression term is dropped in \eq{rg} such that
\begin{align}
 \frac{R_g}{\rsmf} &\simeq \left(\frac{\effvol}{\vsm}\right)^{1/5} \nsm^{3/5}.
 \label{good}
\end{align}
In this \emph{good solvent} regime, the monomers form a self-avoiding random walk. In the limit that $\effvol\rightarrow \vsm$ (which corresponds here to an absence of depletants), the good solvent regime concludes with the extreme \emph{athermal solvent} $R_\textmd{athermal} \simeq \nsm^{3/5}\rsmf$.

\paragraph{Poor Solvent}
\label{poorSolvent}
If the chain is in the collapsed state then $R_g \ll \rsmf N^{1/2}$ so $F_{ent}\approx0$ and only the two interaction free energy terms remain such that
\begin{align}
 \frac{R_g}{\rsmf} &\simeq \left(-\frac{\omega}{\vsm \effvol }\right)^{1/3} \nsm^{1/3}, 
\end{align}
where the negative sign within the brackets is appropriate since $\effvol$ is expected to be negative. 
This $\nsm^{1/3}$ scaling is what we would expect for a polymer in a \emph{poor solvent}.

The poor solvent collapse can only endure for so long. 
Eventually the polymer is in its fully globular state and $R_\textmd{glob} \simeq \nsm^{1/3}\rsmf$, 
which, of course, scales the same as the poor-solvent case but no longer varies with increased $\phidep$ through $\effvol$. 
This extrema is referred to as a \emph{non-solvent}. 

\paragraph{Theta-Solvent}
\label{thetaSolvent}
The analogy of solvent quality as a framework for discussing entropic effects of depletants suggests there will exist some depletant volume fraction that corresponds to a theta-solvent condition. 
This will be denoted $\phidep^\Theta$. 
In this situation, the theta-solvent condition is controlled by the volume fraction of depletants instead of temperature as in a traditional solvent. 
The theta-point exists between good solvent and poor solvent conditions and occurs when the radius of gyration scales as an ideal random walk
\begin{align}
 \frac{R_{g}}{\rsmf} &\simeq \nsm^{1/2}.
 \label{ideal}
\end{align}
This suggests that $\effvol\simeq0$. 
Inserting $R_{g} \simeq \rsmf \nsm^{1/2}$ into \eq{rg}, we find that $\effvol \simeq -\left(\omega/{\rsmf}^3\right) \nsm^{-1/2} \propto \nsm^{-1/2}$, which only goes to zero in the limit of long chromosome chains. 
For the rather short chains of length $\nsm=15$ used in this study, we expect the theta-regime to exist over a narrow range of depletant volume fractions. 
\end{subequations}

%%%%%%%%%%%%%%%%%%%%%%%%%%%%%%%%%%%%%%%%%%%%%%%%%%%%%%%%%%%%
%%%%%%%%%%%%%%%%%%%%%%%%%%%%%%%%%%%%%%%%%%%%%%%%%%%%%%%%%%%%
%%%%%%%%%%%%%%%%%%%%%%%%%%%%%%%%%%%%%%%%%%%%%%%%%%%%%%%%%%%%
%Phase transition
%%%%%%%%%%%%%%%%%%%%%%%%%%%%%%%%%%%%%%%%%%%%%%%%%%%%%%%%%%%%
%%%%%%%%%%%%%%%%%%%%%%%%%%%%%%%%%%%%%%%%%%%%%%%%%%%%%%%%%%%%
%%%%%%%%%%%%%%%%%%%%%%%%%%%%%%%%%%%%%%%%%%%%%%%%%%%%%%%%%%%%
\section{Coil-Globule Collapse}
\label{sctn:phaseTrans}

In the previous section, we described the good and poor solvent regimes (with a narrow theta-solvent regime between them) that are predicted from the Flory free energy framework. 
This is a description that is general to polymer physics. 
Making it specific to the model chromosome in a bath of inert, depletant proteins requires that the effective volume and three-body coefficient of the structural monomers be determined. 
Once the effective volume and three-body term are approximated, the radius of gyration of the chromosome chain can be calculated as a function of $\phidep$ from \eq{rg}. 
In our simplified expression for the free energy, the three-body term is assumed to be constant and approximated by the hard-sphere value
\begin{align}
 \threebody &\approx \frac{\bsm{3}}{10} \approx {\vsm}^2,
\end{align}
since triplet interactions are small~\cite{goulding01,zhu03,ashton14}. 
The second virial coefficient $\bsm{2}$ provides the effective volume of the structural monomers $\effvol \equiv \bsm{2}/4 $. 

\begin{figure}[tb]
 \begin{center}
  \includegraphics[width=0.49\textwidth]{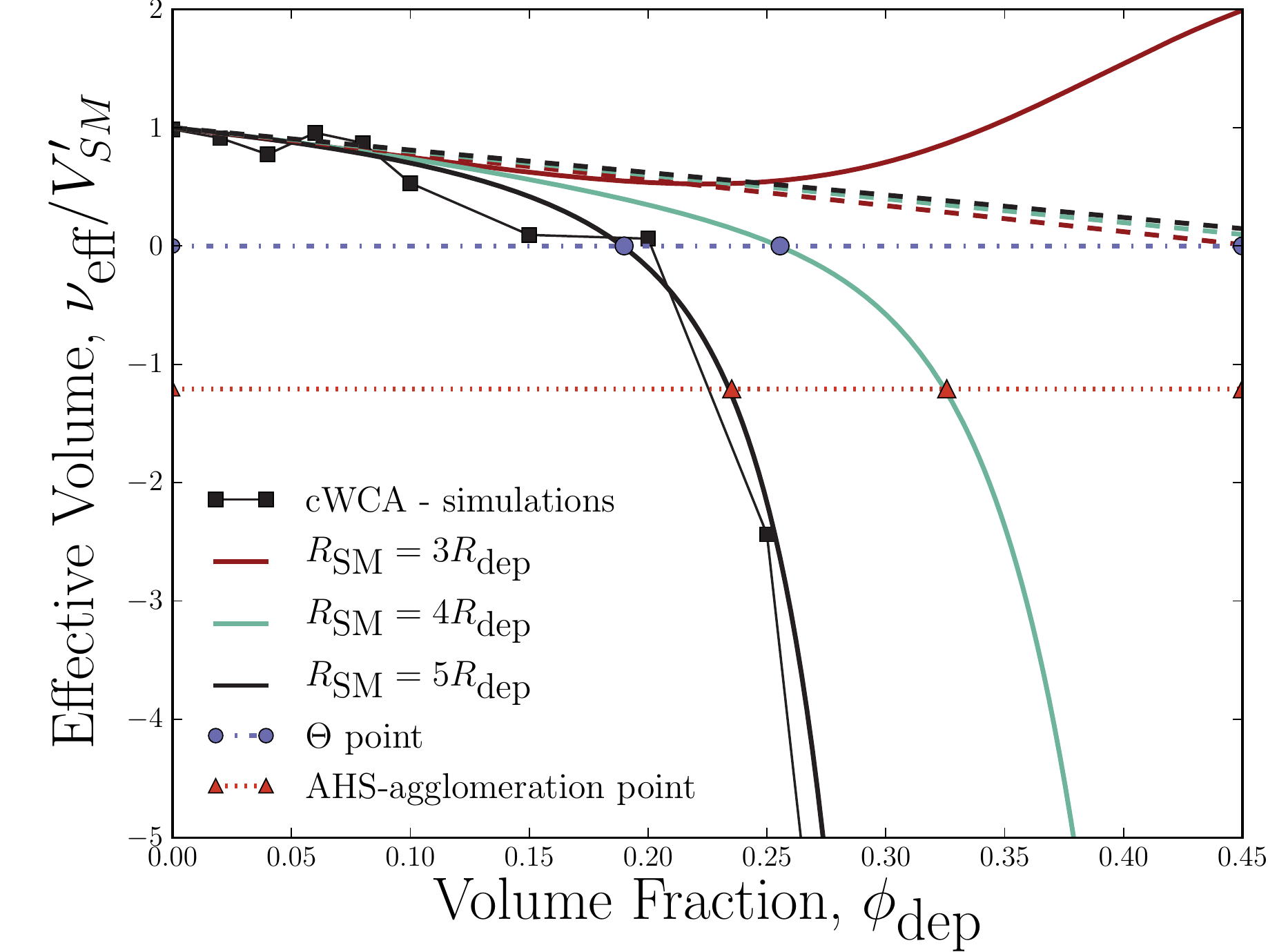}
  \caption{
  The effective volume of structural monomers interacting via depletant-induced pair potentials as a function of depletant volume fraction $\phidep$. 
  Simulation results for the size ratio of $\rsm/\rdep=5$ (black squares) compare well with both the linear Asakura-Oosawa (dashed lines) and the Morphometric Thermodynamics (solid lines) models at low volume fractions, approaching the physical volume $\vsm$ as $\phidep\rightarrow0$. 
  At higher volume fractions the simulations and MT model drop rapidly to negative values, passing through the theta-point $\effvol^\Theta$ (blue dotted line and circles) and the AHS-agglomeration point $\effvol^{*_\textmd{AHS}}$ (red dash-dot line and triangles).
  }
  \label{fig:effVol}
 \end{center}
\end{figure}

The effective volume of the structural monomers is calculated from the total interaction energy $\utot$ between structural monomers via the second virial coefficient \ie
\begin{align}
 \effvol &= -\frac{\pi}{2}\int_0^\infty \left( e^{-\utot/\kbt} - 1 \right) r^2 dr .
 \label{effvol}
\end{align}
The structural monomers interact through two pair-potentials. 
The first is the repulsive ``molecular-type'' combinatorial-WCA potential $\uWCA$ given by \eq{eq:WCA}. 
The second interaction is the entropic depletion-induced pair potential $\udep$, such that the total pair potential is the sum 
\begin{align}
 \utot &= \uWCA + \udep. 
 \label{tot}
\end{align}

The depletion-induced pair potential can be estimated in a number of ways. 
Numerical methods are often employed for binary hard-sphere systems~\cite{roth00,schmidt00,roth01,schmidt03,roth99,roth10,oettel04,egorov04,roth06,konig06,herring07,oettel09} and an analytical theory exists for the limit that $\rsm/\rdep\rightarrow1$~\cite{zaccone12} but the conceptually simple framework of Morphometric Thermodynamics (MT)~\cite{oettel09,botan09,ashton11} is found to be sufficient for the size ratios considered here. 
The MT model is able to reproduce up to the first repulsive/anti-correlation component of the pair potential in agreement with simulations~\cite{part1}. 
It does so by modelling the entropic interaction as arising from changes to the accessible volume $V_o$ with its conjugate osmotic pressure $\Pi$, to the surface area $A$ restriction with the resulting entropic surface tension $\gamma$, and finally to the Gaussian curvatures $C_1$ and $C_2$ with corresponding entropic bending rigidities $\kappa_1$ and $\kappa_2$,
\begin{align}
 \udep\equiv \umtm &\approx \Pi V_o + \gamma A + \kappa_1 C_{1} + \kappa_2 C_{2} .
 \label{morpho}
\end{align}
The geometric coefficients are simple functions of separation and the appropriate effective colloid size
for cWCA spheres as calculated from the second virial coefficient in the absence of depletants $\rsmf$. However, the thermodynamic quantities are less straightforward, though they can be found in the literature~\cite{rosenfeld89,botan09}. 
The MT model is an improvement over the Asakura-Oosawa (AO) model for dilute systems~\cite{asakura54,lekkerkerker}. 
The second virial coefficients and depletion-induced pair potentials between WCA colloids have been demonstrated to be well approximated by the MT model~\cite{part1}. 

Substituting \eq{morpho} and \eq{tot} into \eq{effvol} and \eq{solventQuality} predicts a solvent quality through an effective excluded volume of the structural monomers that is well approximated as $\vsm$ at low volume fractions of depletants but drops rapidly to large negative numbers at higher $\phidep$ values just as the $\rsm/\rdep=5$ simulations produce (\fig{fig:effVol}). 
This is in contrast to the AO model, which, although surprisingly accurate for volume fractions $\phidep\lesssim0.15$, continues to predict a linear decrease even for large volume fractions (\fig{fig:effVol}). 
For $\rsm/\rdep=4$, the predicted behaviour of the effective volume curve as a function of depletant volume fraction is qualitatively similar to the $\rsm/\rdep=5$ at these volume fractions --- at low volume fractions the effective volume is well predicted by the AO model and at higher volume fractions $\effvol$ drops rapidly to large negative numbers.
However, the $\rsm/\rdep=3$ curve is quite different. 
Rather than dropping at large volume fractions, $\effvol$ begins to climb back up for $\rsm/\rdep=3$ (\fig{fig:effVol}). 
This is a characteristic behaviour of the MT model for small size ratios and high volume fractions of depletants~\cite{ashton11} and demonstrates that artificial and nonphysical artifacts dominate $\effvol$ in this limit. 
In the large $\phidep$ and near-unity $\rsm/\rdep$ limit, the MT model generally over predicts the depletion-induced pair potential's primary repulsive barrier. 
The rise of the effective volume curve seen in \fig{fig:effVol} demonstrates that this occurs at rather small volume fractions for $\rsm/\rdep=3$. 

The MT model can be used to predict theoretically the solvent quality and effective volume $\effvol$ of the structural monomers in a thermal bath of depletants. Using this curve, the Flory formalism can estimate the radius of gyration of the model chromosome. 
Only the size ratio $\rsm/\rdep$ and the number of structural monomers $\nsm$ are needed to predict the coil-globule transition. 
In particular, this formalism can estimate the critical volume fraction $\phidep^*$ for the coil-globular collapse. 
When $\rsm/\rdep=5$, the effective volume of the structural monomers crosses zero at $\phidep^\Theta=0.189$ (\fig{fig:effVol}; blue circle). 
Not coincidentally, this is in the range of the $\rg$ collapse in \fig{fig:softCollapse}. 
The same qualitative statement can be said of the $\rsm/\rdep=4$ prediction for the effective volume, except that the predicted theta-point is shifted to higher volume fractions of depletants. 
The steric-repulsion between structural monomers no longer dominates over the depletion-induced attractions at $\phidep^\Theta$ and at higher volume fractions, attractions are more significant causing inevitable collapse. 
Therefore, since the effective volume of the structural monomers is dropping rapidly, the theta-point $\phidep^\Theta$ acts as a rough, lower estimate of the critical point for the coil-globule collapse $\phidep^*$. 

When attractive potentials are short ranged, as is the case with depletion forces, there is a quasi-universality to the critical value at which agglomeration of unconnected hard-spheres occurs~\cite{ashton11}. 
It has been argued~\cite{ashton11} that when $\rsm/\rdep\gg1$, a binary hard sphere mixture of colloids and depletants is accurately equivalent to an ensemble of Adhesive Hard Spheres (AHS)~\cite{miller04}. 
In solutions of AHS, the critical second virial coefficient at which a phase transition to agglomeration occurs is expected to be $\bsm{2}^*=-1.207 \left( 4 \vsm \right)$. 
This leads one to expect that the critical effective volume for which a phase transition is expected $\effvol^{*_\textmd{AHS}}=-1.207\vsm$, rather than $\effvol^\Theta=0$~\cite{ashton11}.  
The AHS-agglomeration point is $\phidep^{*_\textmd{AHS}}=0.234$ when $\rsm/\rdep=5$ (\fig{fig:effVol}; red triangle), which differs from the $\phidep^\Theta$ value by $19\%$. 
In order to encompass these possible metrics for the critical point, we write $\effvol^X \equiv c^X \vsm$ where $X=\left\{*_\textmd{cWCA},*_\textmd{AHS},\Theta\right\}$, $c^\Theta \approx 0$, $c^{*_\textmd{AHS}} \approx -1.207$ and $c^{*_\textmd{cWCA}}$ is measured from the simulations. 

\begin{figure}[tb]
 \begin{center}
  \includegraphics[width=0.49\textwidth]{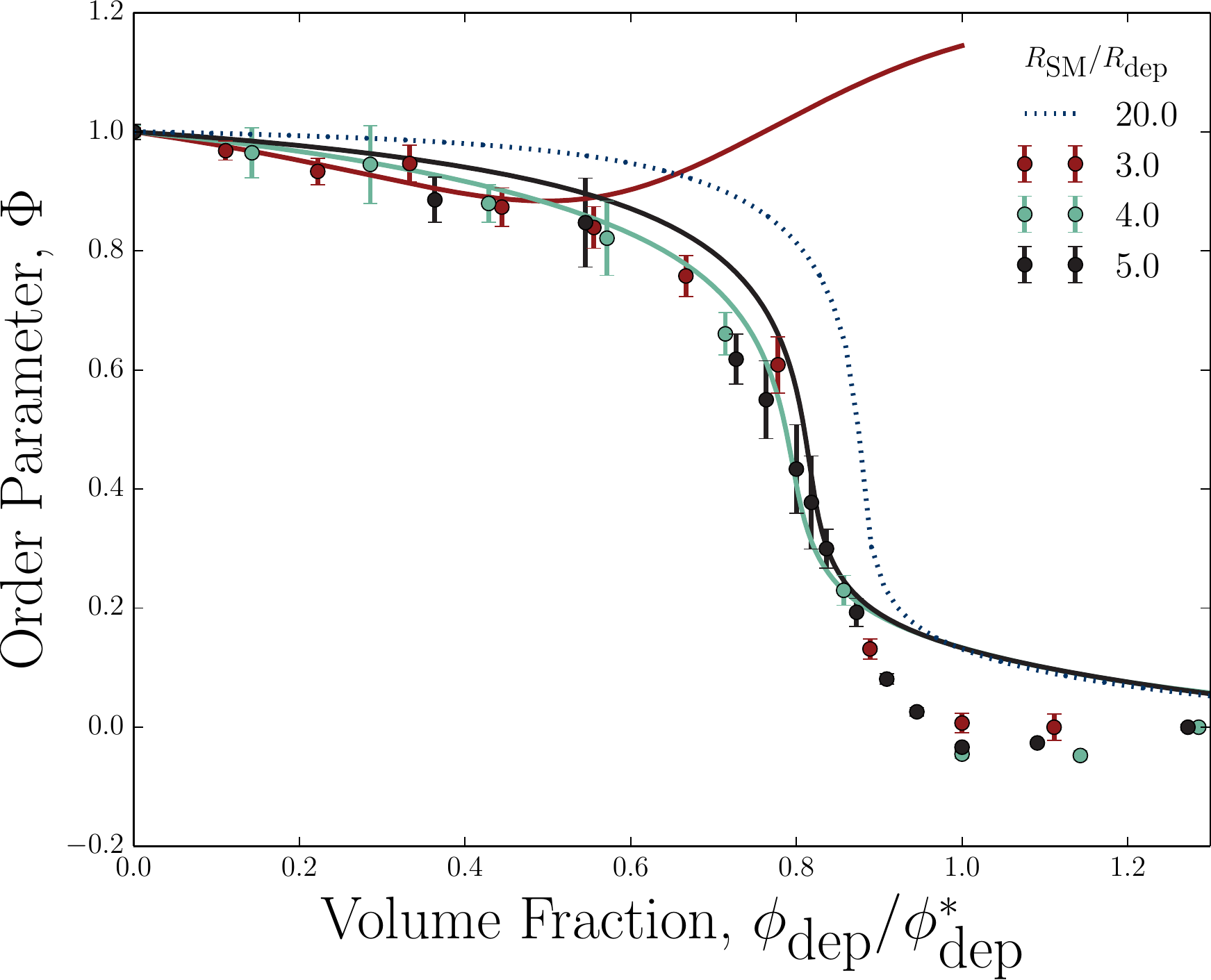}
  \caption{The order parameter $\Phi \equiv \left(\rg - R_\textmd{glob}\right)/\left(R_\textmd{athermal}-R_\textmd{glob} \right)$ reduces the radius of gyration axis of \fig{fig:softCollapse}. 
  When plotted against the rescaled volume fraction $ \phidep/\phidep^{*_\textmd{cWCA}}$ the simulation curves for $\rsm/\rdep=\left\{3,4,5\right\}$ collapse to a single curve. 
  The theory results from substituting the effective volume from \eq{effvol} into the Flory theory (\eq{rg}) and estimating the AHS-agglomeration point $\phidep^{*_\textmd{AHS}}$ as the critical volume fraction. 
  In addition to the size ratios simulated, the Flory prediction for a larger size ratio of $\rsm/\rdep=20$ is also included as a dashed curve. 
  }
  \label{fig:fit}
 \end{center}
\end{figure}

The critical point can be used to rescale the volume fraction of depletants and an order parameter for the coil-globule collapse can be defined. 
This coil-globule order parameter is 
\begin{align}
 \Phi &\equiv \frac{\rg - R_\textmd{glob}}{R_\textmd{athermal}-R_\textmd{glob} } .
\end{align}
For the cWCA simulations, $R_\textmd{athermal}$ is taken to be the $\phidep=0$ value of $\rg$ and $R_\textmd{glob}$ the radius of gyration at the largest volume fraction considered. 
When the order parameter $\Phi$ is plotted against the rescaled volume fraction $\phidep/\phidep^{*_\textmd{cWCA}}$, the three explicit combinatorial-WCA simulation curves from \fig{fig:softCollapse} collapse onto a single curve (\fig{fig:fit}). 
The measured order parameter starts at unity when the number of depletants is zero, decreases slowly over low depletant volume fractions and transitions to zero at the critical point $\phidep/\phidep^{*_\textmd{cWCA}}=1$. 
For higher volume fractions the order parameter remains $\Phi=0$. 

Likewise, the theoretical order parameter curve can be predicted by substituting the predicted effective volume for the pair potential (\eq{effvol}) into the Flory theory (\eq{rg}) to determine the radius of gyration as a function of depletant volume fraction (\fig{fig:fit}). 
The predicted radii of gyration are well represented by this implicit Flory theory when $\rsm/\rdep=5$ and $4$. 
The curves nearly collapse and show good agreement with the simulations when the volume fraction of depletants is scaled by the AHS agglomeration point $\phidep/\phidep^{*_\textmd{AHS}}$. 
The low volume fraction region of negative curvature is the well-predicted region while simulations appear to begin to collapse slightly sooner. 
This causes the radius of gyration to be over-predicted during the transition. 
\fig{fig:fit} is particularly remarkable because of the simplicity of the Flory theory used. 
However, because the MT model fails to predict the effective volume when $\rsm/\rdep=3$, the resulting Flory theory is inadequate as well for such a near-unity size ratio. 
On the other hand, the implicit Flory theory using the effective structural monomer volume from the MT approximation is well behaved for large, more realistic size ratios of $\rsm/\rdep \gtrsim 20$. 
One such theoretical transition is shown in \fig{fig:fit} (blue dashed curve). 
For such large size ratios the coil-globule collapse is indeed steeper but does not become discontinuous. 
When considering the depletion-induced solvent quality for such large size ratios, one must use prudence: 
because the coarse-grained simulations presented here model the structural monomers as impenetrable beads without internal conformational entropy or spaces into which small proteins may translate, the depletion-induced well depth continues to deepen and the critical $\phidep^*$ progressively decreases as the size ratio increases. 
By $\rsm/\rdep = 20$, the critical volume fraction for the coil-globular transition has shrunk to $\phidep^{*_\textmd{AHS}}=0.029$. 

The coil-globule transition of polymer chains in implicit and explicit solvents has been considered extensively for decades~\cite{eizner69,post79,williams81,birshtein91,grosberg92,polson05} and the transition order has been well characterized~\cite{yang13}. 
The Flory theory is able to predict either a first-order or second-order phase transition and indeed has been used to account for the first-order condensation of DNA in the presence of trivalent spermidine~\cite{yoshikawa96}. 
For an implicit solvent, the third virial coefficient controls the nature of the transition and the critical value is $\critThree/\rsm^6 \sim 10^{-2} $~\cite{yang13,grosberg97}. 
Below $\critThree$ the transition is first-order, while the behaviour is a continuous ``crossover'' transition for $\threebody > \critThree$~\cite{yang13}. 
For a freely jointed chain of structural monomers interacting via non-additive depletion-induced pair potentials, the three-body interaction coefficient is comparable to ${\vsm}^2\sim{\rsmf}^6$ and so a crossover transition is expected. 

\section{Discussion}

We have seen that a simple model of bacterial chromosomes as linear chains of freely jointed excluded volume structural monomers is sufficient to account for a continuous coil-globular collapse in the absence of confinement. 
Richer interactions must be significant for \textit{in vivo} bacterial chromosomes to exhibit a first-order phase transition as reported by Pelletier \etal for \textit{in vitro} chromosomes in the presence of inert PEG depletants.~\cite{pelletier12}. 
This difference suggests that mechanisms beyond depletion-induced attraction between freely jointed structural monomers remain substantial in the experimental PEG-chromosome solutions. 

Our results demonstrate that, although depletion effects are sufficient to collapse the chromosome, they are not able to capture the reported first-order nature of the transition. The simplest way to transform the collapse to first-order is through the addition of chain stiffness, but this would most likely be an over simplification. 
Indeed, it is well known that persistence length decreases $\threebody$ and the traditional example of a polymer possessing a first-order coil-globule phase transition is a semi-flexible chain in poor solvent~\cite{grosberg94,grosberg97}. 
In the case of bacterial chromosomes, the structural monomers may not be well approximated as spherical structural monomers. 
Additionally, physical complications such as electrostatic, dielectrophoretic and hydrophobic interactions may modify three-body interactions. 

While stiffness may play a role, it is far more likely that biologically significant interactions between proteins and DNA alter the nature of chromosome condensation. 
Nucleoid-associated proteins acting enthalpically between structural monomers may significantly affect the three-body interactions, which could alter the nature of the collapse and determine globule-state architecture~\cite{mirny11,benza12}. 
There is extensive evidence in the literature that chromosomes form mesoscale structures~\cite{minsky04,stavans06}.
These can only result from specific protein-DNA interactions. 
It is therefore not surprising that coarse-grained computational models of the compaction process must account for these biological associations.
While our work suggests the specific binding mediated by proteins are important, it also reinforces the view that depletion-induced attraction likely plays a significant role in compaction/de-compaction events, as depletion-induced effects may facilitate the action of other enthalpic interactions between DNA and proteins by pre-positioning DNA in these interactions with proteins. 

Our work suggests that the relative importance of generic entropic effects compared to specific protein-DNA interactions can be explored through simple experiments. 
By measuring the critical volume fraction at which collapse occurs in a solution of neutral PEG polymers as a function of the degree of polymerization of the PEG, the relative importance of depletion effects could be found. 
While the critical point for depletion-induced collapse was found here to vary as a function of the size ratio between structural monomers and depletants, collapse mediated by specific protein action should be relatively unaffected by volume fraction of PEG and the critical point is predicted to remain constant. 

Additionally, our simulations could be directly compared to experiments measuring the collapse of chains of microscopic colloids in solutions of colloidal or polymer depletants~\cite{leunissen09,leunissen09b}. 
Such colloidal chains have been synthesized and correspond very closely to the simulations performed here. 
Our results suggest that, unlike chromosomal matter, the colloidal chains will exhibit a continuous coil-globule collapse. 
Such experimental data would provide further evidence that the known enthalpic protein interactions are essential to the nature of chromosome compaction and de-compaction. 

Confinement generally plays an important role and, in the case of confined chromosomal material confinement may be extremely important. 
In relatively small volume fractions of depletants, the chromosome avoids walls to maximize its conformational entropy~\cite{mondal11} but at higher volume fractions depletion-induced attractions to the boundaries may facilitate the coil-globule collapse. 
This competition may have an important impact on the induction of entropic chromosome segregation~\cite{minina14}. 
In particular, the volume fraction of depletants may vary throughout the cell cycle causing compaction-decompaction transitions. 
Likewise, coarse-grained simulations of the sort utilized in this manuscript could illuminate the impact of depletant crowding on the well-known compaction of DNA after replication and perhaps on the unfolding of the chromosome that proceeds replication. 

In order to investigate the possibility of first-order coil-globule collapse and entropic segregation, we propose that future researchers perform simulations of more complex chromosome models and confinement geometries in which depletants are included implicitly rather than explicitly. 
This can be done by including the MT approximation for depletion-induced pair potentials in the interactions between monomers. 
Significantly longer structural monomer chains with narrower good solvent regimes could be simulated to verify that scaling with contour length changes throughout the coil-globular collapse. 
By further including rigidity, we expect that simulations of this nature will observe a discontinuous phase transition from a swollen state to a collapsed globule that passes through metastable states similar to those observed in more traditional solvents~\cite{schnurr00,schnurr02,cooke04,montesi04,lee06,lappala13}. 
In addition, implicit-depletants simulations would have the ability to consider confinement conditions, such as those experienced by intracellular chromosomes and it has been shown that confinement can play a significant role in the collapse of semiflexible chains~\cite{das10}. 

%%%%%%%%%%%%%%%%%%%%%%%%%%%%%%%%%%%%%%%%%%%%%%%%%%%%%%%%%%%%
%%%%%%%%%%%%%%%%%%%%%%%%%%%%%%%%%%%%%%%%%%%%%%%%%%%%%%%%%%%%
%%%%%%%%%%%%%%%%%%%%%%%%%%%%%%%%%%%%%%%%%%%%%%%%%%%%%%%%%%%%
%Conclusions
%%%%%%%%%%%%%%%%%%%%%%%%%%%%%%%%%%%%%%%%%%%%%%%%%%%%%%%%%%%%
%%%%%%%%%%%%%%%%%%%%%%%%%%%%%%%%%%%%%%%%%%%%%%%%%%%%%%%%%%%%
%%%%%%%%%%%%%%%%%%%%%%%%%%%%%%%%%%%%%%%%%%%%%%%%%%%%%%%%%%%%
\section{Concluding Remarks} \label{conclusion}
In this paper we have presented truncated Lennard-Jones (WCA) simulations of coarse-grained bacterial chromosomes in baths of smaller depletant particles in free solution. 
These simulations explicitly demonstrate that the depletant-induced attraction between the chromosome's structural monomers is sufficient to cause the collapse from a swollen state to a globular state. 
The coil-globular collapse was studied as a function of size ratio between structural monomers and depletants, which may have implications for a variety of biopolymers in crowded environments. 
We demonstrate that, within these simulations, the coil-globular collapse is a crossover transition analogous to what one would expect for a freely jointed polymer chain transitioning from good-solvent conditions to poor solvent. 

The effective solvent quality is quantified by predicting the effective excluded volume of each structural monomer. 
We propose that the effective volume can be well approximated for combinatorial-WCA simulations by modelling the total pair interaction as the WCA potential plus the Morphometric Thermodynamics (MT) model for the depletant-induced pair potential. 
For sufficiently large ratios of structural monomer to depletant size, the MT model accurately predicts the effective volume of the structural monomers. 
Through this effective volume, the critical volume fraction can be estimated and we find that the theoretical prediction of the radius of gyration as a function of volume fraction of depletants agrees with the coil-globule collapse observed in the simulations. 
Both simulations and theory for this simplified model of bacterial chromosomes in a bath of protein depletants predict that depletant-induced attractions are sufficient to cause a continuous collapse to a globular state. 
In order to have a first-order phase transition as reported experimentally, our explicit simulations demonstrate that further physical features such as confinement effects, more complicated structural monomers or enthalpic effects due to specific protein-DNA interactions must be included. 

\section*{Acknowledgements}
We gratefully acknowledge support through NSERC Discovery Grants to G.W.S. and J.L.H. and EMBO funding to T.N.S (ALTF181-2013).  
High performance computational resources were graciously provided by Prof. Laura Ramuno at the University of Ottawa and Sharcnet. We are thankful for helpful suggestions from Tina Haase and Owen Hickey. 

\section*{References}
\bibliographystyle{unsrt}
\bibliography{depletantSources}

\end{document}